\documentclass[conference]{IEEEtran}
\usepackage{nopageno}

\ifCLASSINFOpdf
\else
\fi

\usepackage[colorlinks, citecolor=blue, bookmarks=false]{hyperref}
\usepackage{setspace, amsmath, amssymb, url, lscape, subfigure, algorithmic, multirow, pslatex, verbatim, alltt, amsfonts, wrapfig, color, cite} 
\usepackage[vlined,linesnumbered,ruled,boxed]{algorithm2e}

\usepackage{balance}
\usepackage[dvips]{graphicx}
\usepackage{epsfig}

\usepackage{xcolor}
\newcommand{\qed}{\nobreak \ifvmode \relax \else
      \ifdim\lastskip<1.5em \hskip-\lastskip
     \hskip1.5em plus0em minus0.5em \fi \nobreak
      \vrule height0.75em width0.5em depth0.25em\fi}

\newcommand{\bs}{\text{Base Station}}
\newcommand{\lb}{\text{Load Balancer}}
\newcommand{\eg}{{\it e.g.,}}

\newcommand{\ie}{{\it i.e.,}}

\newcommand{\comments}[1]{}
\newcommand\hl{\bgroup\markoverwith
  {\textcolor{yellow}{\rule[-.5ex]{2pt}{2.5ex}}}\ULon}

\begin{document}

\title{Robust Resource Allocation Using Edge Computing for Vehicle to Infrastructure (V2I) Networks}

\author{
	{\bfseries Anna Kovalenko$^1$, Razin Farhan Hussain$^1$, Omid Semiari$^2$, and Mohsen Amini Salehi$^1$}\\
	$^1${\{aok8889, razinfarhan.hussain1, amini\}@louisiana.edu} $^2${osemiari@georgiasouthern.edu}\\
	$^1$High Performance Cloud Computing (\href{http://hpcclab.org/}{HPCC}) Laboratory,\\ School of Computing and Informatics, University of Louisiana at Lafayette, Lafayette, LA, USA\\
	$^2$Department of Electrical and Computer Engineering, Georgia Southern University, Statesboro, GA, USA\\
}

\maketitle
\thispagestyle{empty}
\IEEEpeerreviewmaketitle
\begin{abstract}
Development of autonomous and self-driving vehicles requires agile and reliable services to manage hazardous road situations. Vehicular Network is the medium that can provide high-quality services for self-driving vehicles. The majority of service requests in Vehicular Networks are delay intolerant (\eg \ hazard alerts, lane change warning) and require immediate service. Therefore, Vehicular Networks, and particularly, Vehicle-to-Infrastructure (V2I) systems must provide a consistent real-time response to autonomous vehicles. During peak hours or disasters, when a surge of requests arrives at a \bs, it is challenging for the V2I system to maintain its performance, which can lead to hazardous consequences. Hence, the goal of this research is to develop a V2I system that is robust against uncertain request arrivals. To achieve this goal, we propose to dynamically allocate service requests among \bs s. We develop an uncertainty-aware resource allocation method for the federated environment that assigns arriving requests to a \bs~so that the likelihood of completing it on-time is maximized. We evaluate the system under various workload conditions and oversubscription levels. Simulation results show that edge federation can improve robustness of the V2I system by reducing the overall service miss rate by up to 45\%.
\end{abstract}


\section{Introduction}\label{sec:intro}
\begin{figure}[h!]
\centering
\includegraphics[scale=0.35]{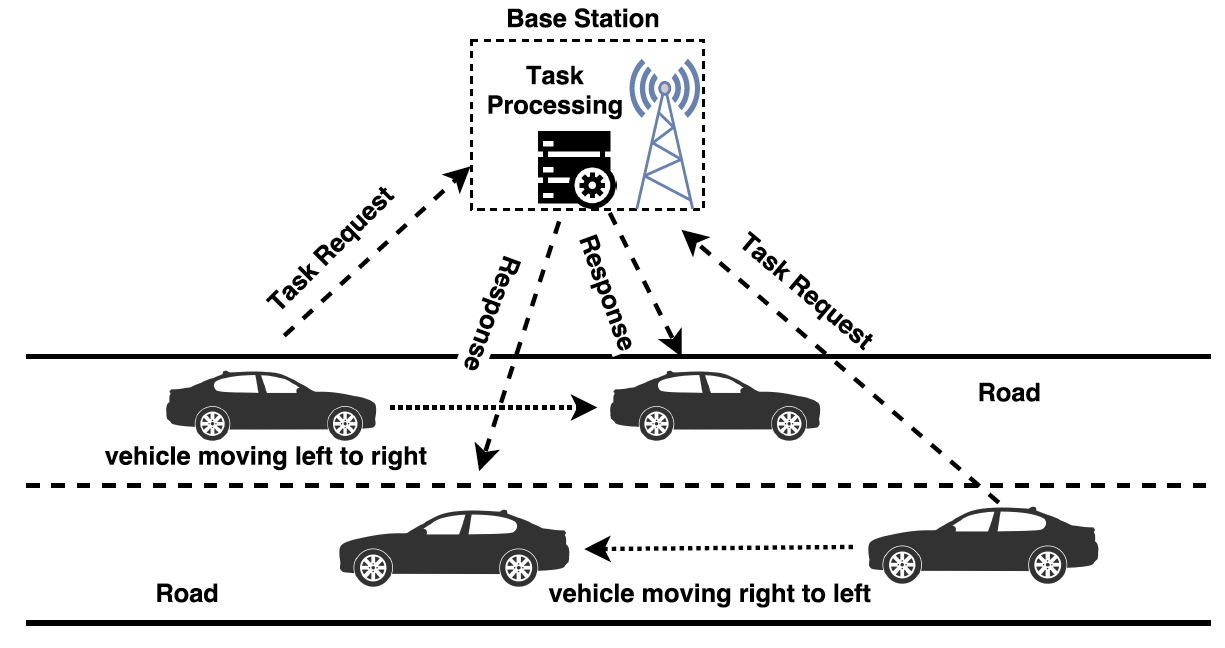}
\caption{A Vehicle to Infrastructure (V2I) scenario where vehicles send requests to a Base Station and receive the response. A Base Station is a roadside unit with communication and computation abilities.\label{fig:scen}}
\end{figure}

Recent advancements in communication and computation technologies have stimulated a rapid development of vehicular networks. Federal Communications Commission (FCC)~\cite{ali2011co} has reserved 5.850 to 5.925 GHz frequency band for Vehicle-to-Everything (\emph{V2X}) communications. Vehicle-to-Infrastructure (V2I) communication is one prominent form of V2X which is visioned under the hood of Intelligent Transport System (ITS) to improve roadside safety and traffic systems. In V2I, infrastructure refers to all the devices (known as Base Stations) that facilitate communications and computations to serve vehicular requests. As shown in Figure~\ref{fig:scen}, autonomous vehicles send their service requests (tasks) to \bs s while operating on the road. A \bs~(BS) is capable of communicating with vehicles and processing vehicular tasks. Examples of such vehicular tasks can be wrong way driver warning~\cite{ABIresearch}, cooperative forward collision warning~\cite{ElBatt}, and lane change warning~\cite{ABIresearch}. This type of tasks can only tolerate a short end-to-end delay, \ie the delay from issuing a task request until receiving its result~\cite{ali2011co}. For such tasks, there is no value in executing them after a tolerable delay.

Owing to the centralized nature of clouds, both Vehicular Cloud Computing (VCC\cite{gerla2012}) and conventional cloud systems incur high latency~\cite{li2017resource}. Nonetheless, \bs's computational power can be harnessed with an edge computing system that can manage vehicular services at the \bs~with low latency without communicating to cloud datacenter. Various \bs s in a V2I system can potentially be heterogeneous, both in terms of computational characteristics and communication medium to the core network (\eg wireless, optical, and wired).
A problem arises during road emergencies (\eg road accidents or disasters) when a rapid increase in service requests to \bs s significantly affects the tasks' service time. In fact, in this situation, \bs~resources become oversubscribed, and many tasks miss their deadlines due to insufficient computational resources.

\emph{Robustness} is defined as the degree to which a V2I system can maintain a certain level of performance even in the presence of stochastic parameters in the system~\cite{jpdcmohsen,li2017cost,ali2004measuring}. Accordingly, our goal, in this research, is to design a V2I system that is robust against uncertain task arrival. In this research, we evaluate robustness of the V2I system according to the number of tasks that can meet their deadlines. The main question we try to answer is stated as how to allocate arriving tasks among the \bs s so, the systems' robustness is maximized? An efficient solution to this problem needs to overcome the uncertainties (\eg uncertain task arrival, task execution times, communication delay) of the system. 
Previous research works either discard these uncertainties\cite{bok2016multiple} or focus only on one of the aforementioned uncertainty factors (\eg communication\cite{ali2011co}). Alternatively, we propose a robust model that copes with uncertainties introduced by the task arrival, communication, and computation. To cope with the uncertainty in task arrival, we propose to federate \bs s and devise a \lb~at the \bs~that can leverage the computational capabilities of neighboring \bs s to improve robustness of the V2I system. 


Allocating vehicular tasks in a V2I system is proven to be an NP-complete problem~\cite{Ullman1975}. Therefore, a large body of research works have been dedicated to developing resource allocation heuristics in such systems~\cite{ali2004measuring,yu2016optimal,li2017resource,liu2010rsu}. As such, we leverage our proposed probabilistic theory and establish a novel resource allocation heuristic for the \bs s' \lb. Then, we evaluate the performance of our proposed heuristic under various workload conditions.
The main contributions of this paper are as follows:

\begin{itemize}
    \item Propose a system that encompasses the uncertainties that exist in task arrival, communication, and computation.
    \item Devise a probabilistic theory to predict the success of task completion within the deadline on different \bs s.
    \item Develop a load balancing heuristic that functions based on the proposed theory and increases the robustness of the V2I system.
    \item Analyze the performance of our proposed system under various workload conditions and in comparison to different existing models.
\end{itemize}

The rest of the paper is organized as follows. Section \ref{sec:scen} introduces the System Model and Problem Statement. Section \ref{sec:approach} discusses the robust V2I system based on federated \bs s. Sections \ref{sec:heuristcic} and \ref{sec:simulation} present resource allocation heuristics and experimental setup respectively. Performance evaluation is presented in section \ref{experiments}. Section \ref{relatedwork} presents Related Work. Finally, section \ref{sec:conclsn} concludes the paper.

\section{System Model and Problem Statement}\label{sec:scen}
\subsubsection{Scenario}

The Road Side Unit, or as we call it, a \bs~is located at the intersection or the side of the road. \bs s are stationary edge devices with memory, storage, computational ability, and short wireless range transmission system \cite{liu2010rsu}. The receiving \bs~is connected to a number of nearby heterogeneous \bs~via 5G wireless network~\cite{yu2016optimal}. In summary, the receiving \bs~can transfer tasks to the neighboring \bs s, that forms a federation of \bs s. Every \bs~in the federation is also connected to a central cloud infrastructure. In this scenario, the roads and intersections are considered to be congested. Therefore, the \bs~situated at the roadsides and intersections is oversubscribed. 
In our scenario, all tasks have an individual deadline, the time frame in which tasks should be processed and returned to the vehicle. Based on the specific service required, tasks are separated into two groups, urgent (delay-intolerant) and non-urgent (delay-tolerant). For example, Hazards Around the Area alerts are considered to be urgent, and the On-board Entertainment updates are considered non-urgent. According to the task type, non-urgent tasks usually are bigger in data size and need more time to execute. Respectively, urgent tasks are smaller in data size. 


\subsubsection{Assumptions}
According to the described scenario, a set of tasks is generated by vehicles and sent to the \bs~for processing. Every task $i$ has its own deadline (denoted $\delta_i$) within which it has to be completed. The V2I system allocates the tasks to \bs s by considering the individual deadline for each task. 
To enable robustness, the resource allocation method aims to maximize the number of tasks meeting their deadlines. 

According to the problem definition, the set of arriving tasks can be defined as $T$, where  $T$ = $\{t_1, t_2, t_3, t_4\dots, t_n\}$ and the set of Base Stations defined as $BS$, where $BS$ = $\{bs_1, bs_2, bs_3, bs_4\dots, bs_m\}$. The set of tasks that can meet their deadlines, denoted $T_s$, is the subset of $T$ ($T_s \subseteq T$).


%
%
%



Upon arrival to a \bs, the task is assigned an individual deadline that includes task arrival time and the end-to-end delay the task can tolerate. We assume in our case that communication delay (uplink and downlink delay) can be significant. Thus, we consider a communication delay for a deadline calculation as well. For arriving task $t_i$, the deadline is defined based on Equation~\eqref{eq:2}.
\vspace{-2mm}
\begin{equation}\label{eq:2} 
\delta_i = arr_i + E_i + \epsilon + \beta 
\end{equation} 

where $arr_\textrm{i}$ is the arrival time of the task, $E_\textrm{i}$ is the average task completion time, $\epsilon$ is a constant value defined by the \bs~(slack time), and $\beta$ is the communication delay. We assume that tasks arrive at the \bs~dynamically, and the arrival rate is not known in advance. As we consider, receiving \bs~is oversubscribed, therefore, some tasks are projected to miss their deadline. If such tasks are delay-sensitive, then they are dropped, which is a common practice in oversubscribed real-time systems \cite{jpdcmohsen}.



\subsubsection{Delay Estimation}
In a V2I system, three distinct factors contribute to the end-to-end delay. They are uplink delay, computational delay, and downlink delay~\cite{mostafa2011}. Therefore, V2I end-to-end delay (denoted $D_\textrm{V2I}$) is defined based on Equation~\eqref{eq:3}. 
\vspace{-2mm}
\begin{equation}\label{eq:3}
D_\textrm{V2I} = d_\textrm{U} + d_\textrm{BS} + d_\textrm{D} 
\end{equation} 

where $d_U$ is average uplink delay, $d_\textrm{BS}$ is average computational delay, and $d_D$ is average downlink delay. 

From Equation \eqref{eq:3}, $d_U$  and $d_D$ can be defined as follows. For task $i$  requested by the vehicle $v$ to the \bs~$m$, the uplink delay from $v$ to $m$ is defined based on Equation~\eqref{eq:4}.
\vspace{-2mm}
\begin{equation}\label{eq:4} 
d_\textrm{U} =  \frac{L_\textrm{i}}{g(v,m)}  
\end{equation} 

and for $t_i$ traveling back from the \bs~$m$ to a vehicle $v$, the downlink delay is defined based on Equation~\eqref{eq:5}.
\vspace{-2mm}
\begin{equation}\label{eq:5} 
d_\textrm{D} =  \frac{L_\textrm{i}}{g(m,v)} 
\end{equation} 

where $L_i$  is the task data size, $g(v,m)$ and $g(m,v)$ is bandwidth for the link from $v$ to $m$ and from $m$ to $v$, respectively.


\section{Robust V2I System based on Federated Base Stations}\label{sec:approach}
\subsection{Federation-Enabled Base Station}
\begin{figure}[h!]
\centering    
\includegraphics[scale=0.35]{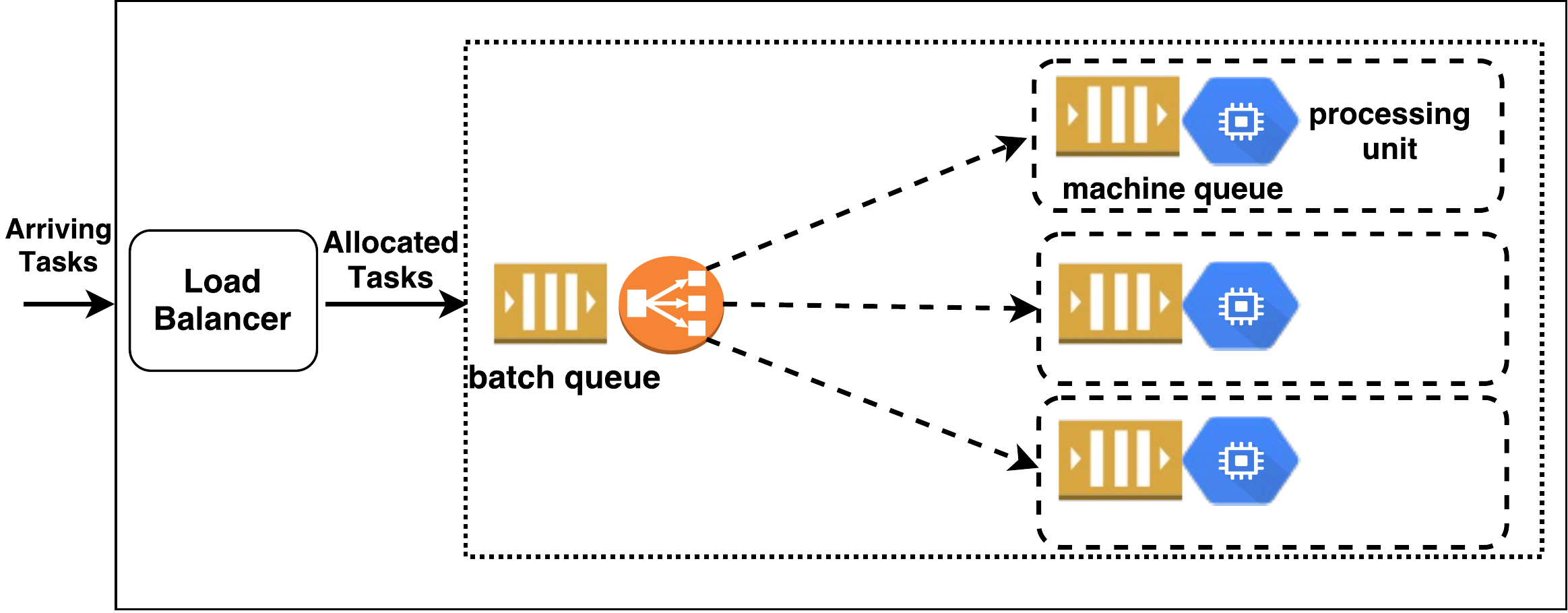}
\caption{Internal structure of the proposed Base Station.}\label{fig:arch}
\end{figure}

To enable the robustness against uncertainty in task request arrival of a V2I system, we propose to federate \bs s in a dynamic manner. Thus, \bs s can augment the computational power upon demand and keep the Quality of Service (QoS), even if faced with oversubscription. Hence, we change the structure of \bs, as depicted in Figure~\ref{fig:arch}.

As shown in Figure \ref{fig:arch}, upon arrival of a task, to a \bs, the \lb~immediately allocates the task either to the receiving \bs~or to a neighboring \bs. A queue is present behind \lb~for the tasks that arrive at the same time. There is no re-allocation considered due to the latency overhead produced by additional task transfer (\ie hops). When the task is allocated to a \bs, it enters batch queue of the \bs~for processing. The time span to complete an arriving task within a \bs~is defined as the \emph{computational delay}. Therefore, $d_\textrm{BS}$ in Equation~\ref{eq:4} is defined as $d_{BS} = d_c$, where $d_c$ is the average computational delay.
\vspace{-2mm}
\subsection{Probability of Meeting Deadline in a Federated V2I System}
\begin{figure}[h!]
\centering
    \includegraphics[scale=0.35]{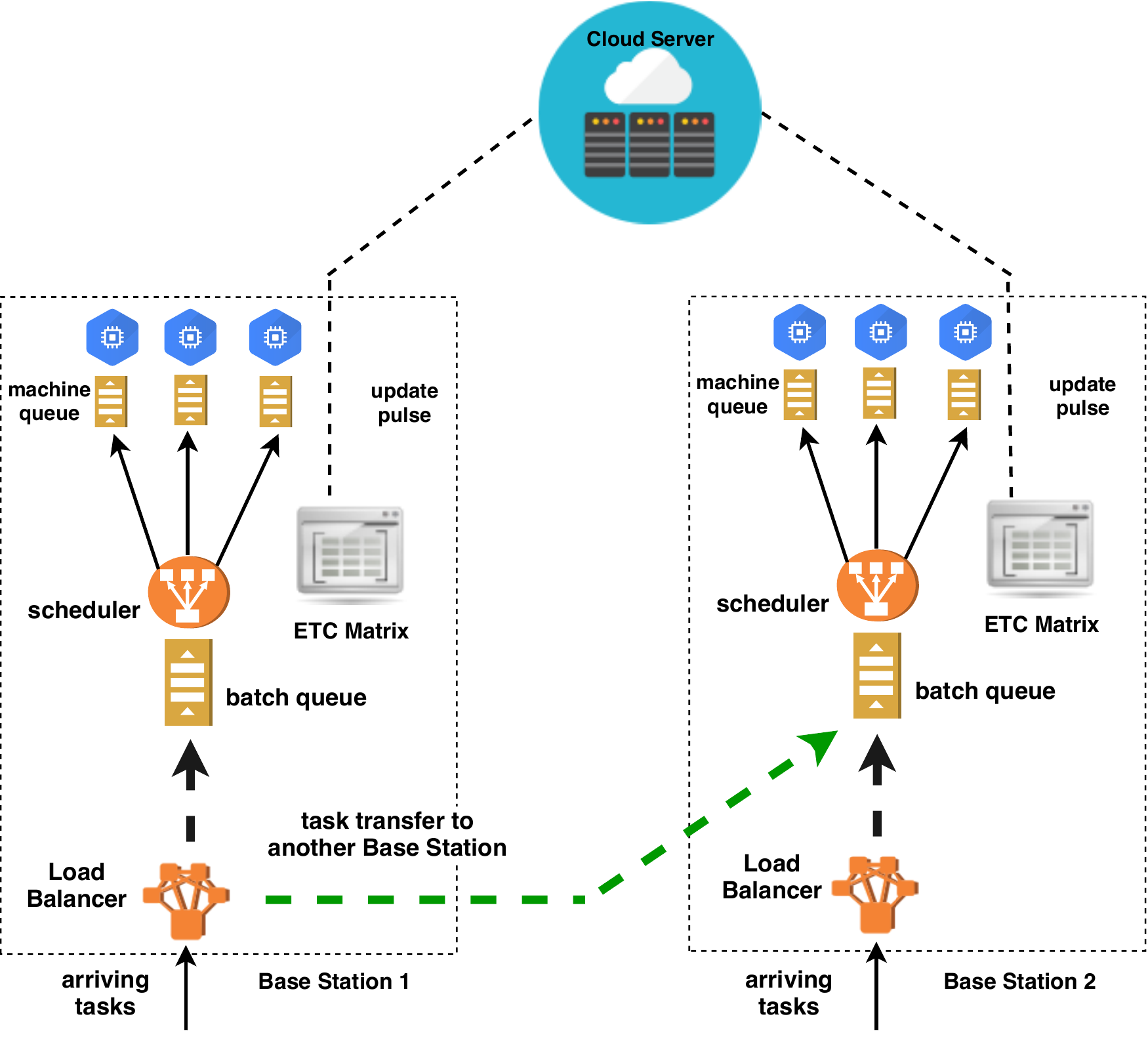}
    \caption{Proposed allocation model for Load Balancer to efficiently allocate arriving tasks in federated edge computing environment.}\label{fig:matrix}
\end{figure}

The \lb~assigns tasks to the \bs~that offers the highest probability of on-time completion (also defined as \emph{task robustness}). 
As shown in Figure~\ref{fig:matrix}, every \bs~maintains two matrices named as Task Completion (ETC) time matrix and Estimated Task Transfer (ETT) time matrix respectively. The ETC Matrix contains estimated task completion time distributions (X$\sim \mathcal{N}(\mu,\,\sigma^{2})$) for different task types in different \bs s. Each ETC matrix entry contains two values, $\mu$ (mean) and $\sigma$ (standard deviation) that are obtained from historical execution time information of each task type on each \bs. We consider that the historic execution time information for each task type on each \bs~form a Normal distribution. In the ETC matrix, every column defines a \bs~and every row defines a task type. 
The Estimated Task Transfer (ETT) time matrix captures the latency of transferring data to another \bs. The entries of ETT are considered as Normal distributions and are obtained from historical data transfer times for different task types from one \bs~to a neighbor. Both ETC and ETT matrices are periodically (\eg approximately every 10 minutes) updated on all \bs s with the help of the central cloud system. 

Using ETC and ETT matrices, \lb~$j$ that receives task $t_i$ (of task type $i$) can calculate the probability of $t_i$ meeting its deadline $\delta_i$ across all \bs s. For the receiving \bs~$j$, the probability of meeting deadline (denoted as $P_i^j$) is defined based on Equation \ref{eq:6}.
\vspace{-2mm}
\begin{equation}\label{eq:6}
P_i^j (\tau(t_i,bs_j) \ \textless \ \delta_i) \ = \ P_i^j(Z \ \textless \ z)
\end{equation}
In Equation~\ref{eq:6}, $\tau(t_i,bs_j)$ is the expected task completion time. The distribution is standardized with $\mu_i= 0$ and $\sigma_i=1$ using the $z$ score that is formulated based on Equation~\ref{eq:7}.
\vspace{-1mm}
\begin{equation}\label{eq:7}
z = \frac{\delta_i \ - \ \mu_i^j}{\sigma_i^j}
\end{equation}


To obtain the probability, we convolve the communication and computational delay distributions to find the overall end-to-end delay distribution. This convolved distribution ($W \sim \mathcal{N}(\mu,\ \sigma^{2}) = X \sim \mathcal{N}(\mu,\ \sigma^{2}) \circledast Y \sim \mathcal{N}(\mu \ \sigma^{2}) $) captures the stochasticity that exists both in communication and computation delays. It is worth noting that, our proposed approach is independent from the type of distribution and can be applied on other distributions too. However, as Normal distribution is more general and calculating convolution is faster, we utilize it in this study. 
Upon receiving a task by the \lb~$j$, for all of its neighboring \bs s, it convolves the  corresponding entry in the ETC matrix with respective entry in the ETT matrix. 
The resulting distribution for each of the neighboring \bs s is used to calculate the probability of the task meeting its deadline on those \bs s. Once the probability is calculated for all of the \bs s, the arriving task is allocated to the \bs~that offers the highest probability. We note that when the task's probability to meet its deadline is zero, the task is dropped.

\section{Resource Allocation Heuristics}\label{sec:heuristcic}
\subsection{Overview}
The following section describes four heuristics considered for the resource allocation. Best Probability is the heuristic we developed based on the proposed model we presented in the previous section. The other two heuristics, Minimum Expected Completion Time, and Maximum Certainty, only utilize the ETC matrix component. Lastly, No Redirection heuristic does not utilize any of the proposed components. 
\vspace{-3mm}
\subsection{Best Probability (BP)}
The Best Probability heuristic considers task robustness as the probability of the specific task to meet its deadline in a particular \bs. 
As shown in Algorithm~\ref{alg1}, when a \lb~receives a task, it first obtains the task's completion time distribution (using ETC) and the transfer time distribution (using ETT) for the received task type across all the \bs s. Next, it convolves two distributions (except for the receiving \bs, where no transfer will occur) to calculate probabilities for the received task to meet its deadline in all neighboring \bs s. Then, the \lb~chooses the \bs~that provides the maximum value (the highest probability) and allocates the task to that \bs.
When two \bs s have the same highest probabilities, then the tie can be resolved by considering the sparsity (known as standard deviation $\sigma$) of two concurrent distributions. Therefore, the preference is given to the distribution with a smaller standard deviation value.

\begin{algorithm}[!h]
	\SetAlgoLined\DontPrintSemicolon
	\SetKwInOut{Input}{Input}
	\SetKwInOut{Output}{Output}
	\SetKwFunction{algo}{algo}
	\SetKwFunction{proc}{Procedure}{}{}
	\SetKwFunction{main}{\textbf{TaskAssignment}}
	\Input{Task $t_i$; $ETC$ and $ETT$ matrices; $B$ (set of neighboring \bs s)}
	\Output{Chosen \bs~$j\in B$ to assign $t_i$}
    $p_r(t_i) \gets$ Probability on receiving \bs~$r$ \;
    Provisionally assign $t_i$ to receiving \bs~$r$\;
    \ForEach{\bs~$j \in B$} {
    	$p_j(t_i) \gets$ Probability on neighbor edge $j$ \;
    	\If {$p_j(t_i) > p_r(t_i)$} {	
    		\small{Assign $t_i$ to neighbor \bs~$j$}\;
    		\small{Break}\;
    	}\uElseIf{$p_j(t_i) = p_r(t_i)$}{
    	    
    		\If{$\sigma_j < \sigma_r$}{
    			\small{Assign $t_i$ to neighbor \bs~$j$}\;
    			\small{Break}\;
    			}
        }	
    }
    \uIf{probability of chosen destination is zero}{
        \small{Drop task $t_i$}\;
    }
        
\caption{Task allocation algorithm for load balancer.}\label{alg1}
\end{algorithm}
\vspace{-4mm}
\subsection{Minimum Expected Completion Time (MECT)}

Minimum Expected Completion Time heuristic is well known and widely described in the recent literatures~\cite{jpdcmohsen,hussain1serverless}. For a received task of a particular type, this heuristic utilizes the ETC matrix to calculate the average expected completion time across all the \bs s and selects the \bs~with minimum expected completion time.

\vspace{-3mm}
\subsection{Maximum Certainty (MC)}
Maximum Certainty heuristic (used in~\cite{hussainrobust}) calculates the difference (called certainty ) between the task's deadline and the average completion time for this task type by utilizing ETC matrix. The task is finally assigned to the \bs~that provided the maximum certainty.
\vspace{-2mm}
\subsection{No Redirection (NR)}
No Redirection heuristic does not transfer the task to the neighboring \bs s. Therefore, whenever the arriving task enters the \lb~of a specific \bs, it has to be allocated to that specific \bs. 

\section{Experimental Setup}\label{sec:simulation}
We have used EdgeCloudSim~\cite{sonmez2017edgecloudsim}, a discrete edge simulator to evaluate the performance of our proposed model. Due to computational resource limitation, we implement \bs s in a form of small data centers, with up to 4 cores having computational capacity of 1600 Million Instructions Per Second (MIPS) for each core. All Virtual Machines (VMs) in a \bs~are homogeneous, \ie~same computational power. Nevertheless, \bs s are heterogeneous \ie~have different computational powers. In simulation, we have a total of 15 \bs s, where 8 of them are the 4-core \bs s and 7 are the 2-core \bs s. Each \bs~in has a specified location in an X-Y plane, which is utilized by the Mobility model of EdgeCloudSim to find the closest \bs~to a simulated vehicle. 
We also utilize 4 types of tasks in our simulation where 2 of those types are urgent and the other 2 are non-urgent. Specifically, we implement a Hazard Alert and the Lane Change Warning as the urgent task types (2000-3000 MIPS). The other two types are the On-Board Entertainment and the Fuel Usage Statistics (10000-15000 MIPS).

For the networking model, we specify the Wireless Local Area Bandwidth (WLAN) with 200 MBps. This delay is imposed only when the \lb~transfers tasks to a neighboring \bs~(known as LAN delay) is set to be 2 seconds. By default, the requests are initially offloaded to the nearest \bs, and only then, the load balancer assign receiving task to suitable \bs.
For the workload generation, the tasks are sampled randomly from the Exponential distribution. Each \bs~receives a certain amount of service requests as a receiving \bs, as well as the certain amount of tasks, transferred from the neighboring \bs s. We consider maintaining ETC and ETT matrices in every \bs~and update them in every 10\% of the workload execution. The workloads are seeded and changed from one trial to another.

 


\section{Performance Evaluation}\label{experiments}

The most important evaluation metric is considered to be the number of service requests (tasks) that miss their deadline. We evaluate the performance for the system as federation of \bs s, as well as for an individual \bs.

\subsubsection{The Impact of Oversubscription Level}
To increase the oversubscription level, we increase the number of vehicles that generate service requests, hence, the number of tasks increases. We start initially with 2,000 vehicles passing through the area covered by the network of 15 \bs s in one hour. Each step we increase the traffic by 1,000 vehicles an hour until we reach 7,000 vehicles. In every trial, the simulation is run for 20 times for each one of the state-of-the-art baseline heuristics introduced earlier.

\begin{figure}[h!]
    \centering    
    \includegraphics[scale=0.45]{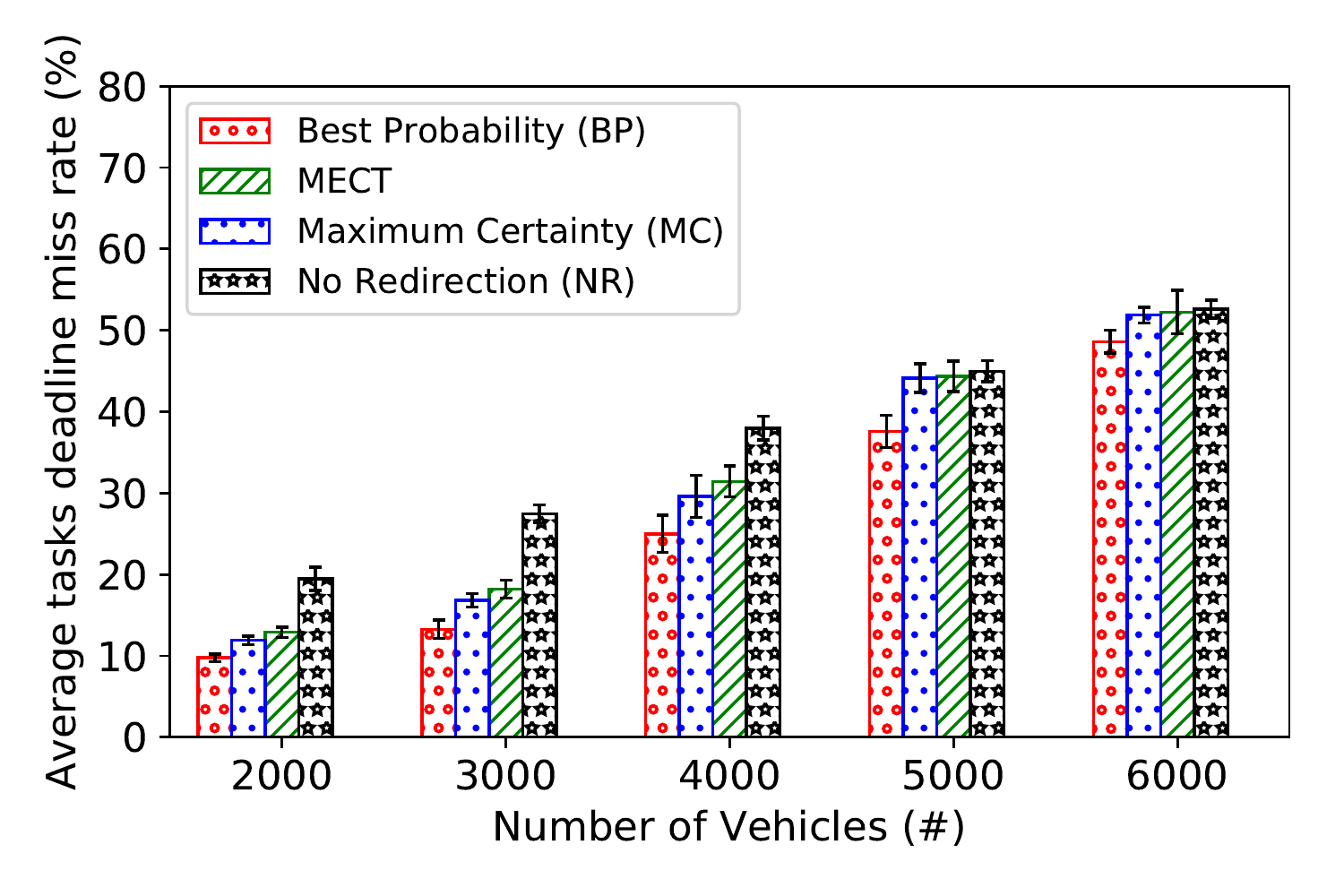}
    \caption{Increasing oversubscription level by increasing the number of vehicles served by the network.}
    \label{fig:oversubscriptionLevel}
\end{figure}
The result of this experiment is shown in Figure~\ref{fig:oversubscriptionLevel}, where the horizontal axis represents the number of vehicles in the system and the vertical axis represents the percent of tasks that missed their deadlines. The number of tasks missing their deadline increases with the higher traffic congestion reflecting the increase in the oversubscription. It also seems reasonable that the No Redirection provides the worst performance compared to all others, as it does not utilizes load balancer.
The Maximum Certainty heuristic outperforms the MECT at every stage by 1-3\%. Finally, we can observe that BP, consistently outperforms other heuristics (3-8\% better then MECT, 4-11\% better than MC, and  4-17\% better than NR) at every stage of the experiment. We can also notice that it shows the best improvement for medium to high levels of oversubscription (\ie \ 3,000---5,000 vehicles). Additionally, even at an extremely high level of oversubscription (\ie \ 6,000 vehicles), when all other heuristics perform nearly the same, our model still shows improvement, demonstrating its robustness against uncertainty in task arrival intensity.

\subsubsection{The Impact of Urgency}
\begin{figure}[h!]
    \centering    
    \includegraphics[scale=0.45]{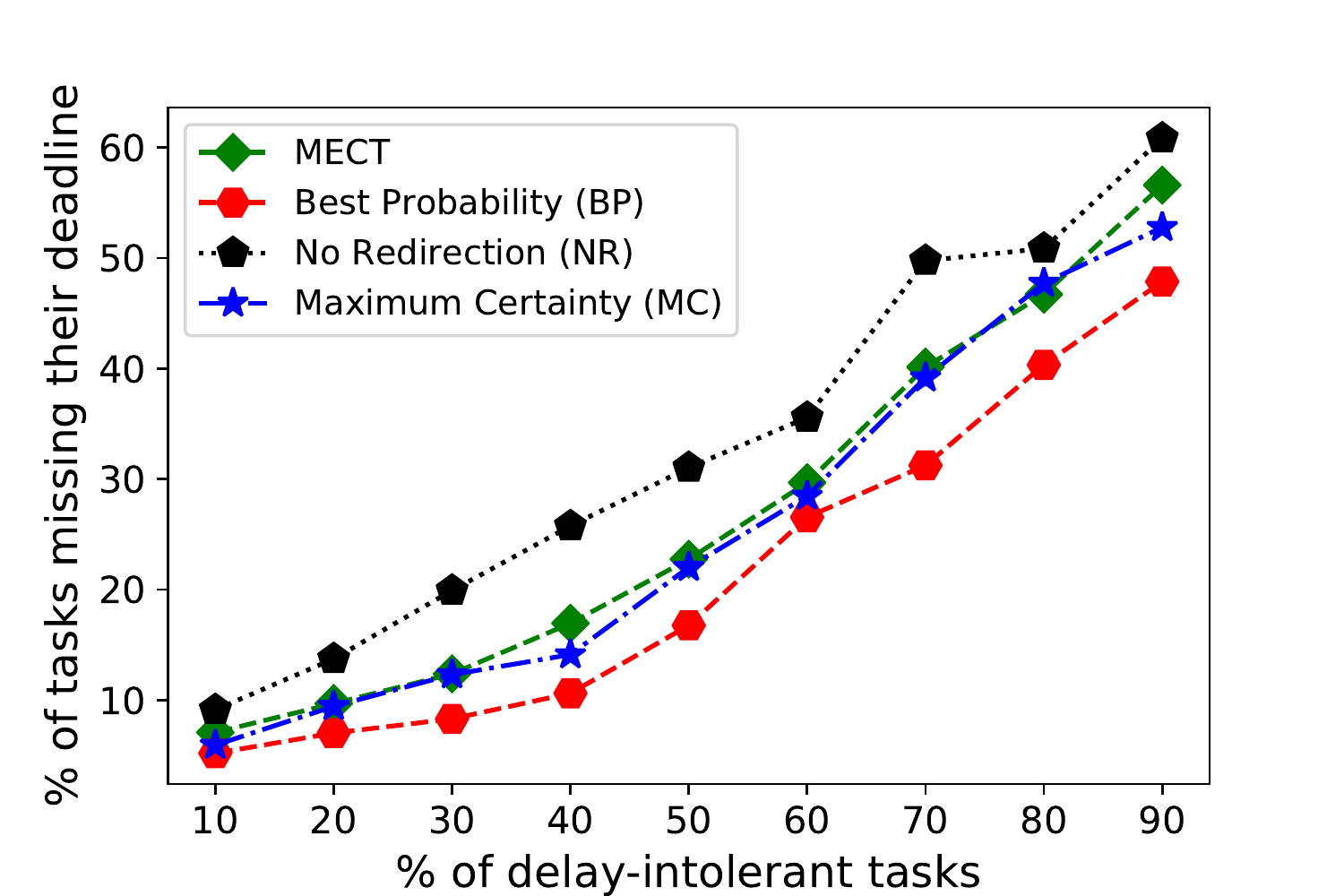}
    \caption{The impact of the portion of urgent and non-urgent service requests in the system.}
    \label{fig:Urgent}
\end{figure}

To evaluate the performance of our proposed model against the task type arrival uncertainty, we increase the ratio of urgent tasks in the workload from 10\% to 90\% (horizontal axis in Figure~\ref{fig:Urgent}) and account for the number of tasks missing their deadline (vertical axis in Figure~\ref{fig:Urgent}) for every heuristic with high oversubscription (4,000 vehicles) level.
Figure~\ref{fig:Urgent} shows that, deadline miss rate rises, as the percentage of urgent tasks increases. The reason is that urgent tasks have a high delay sensitivity compared to non-urgent tasks. Nevertheless, we can see that our heuristic is robust against task type uncertainties. As the No Redirection shows the poorest performance, with MECT and Maximum Certainty outperforming it by 2-10\%, the Best Probability performs consistently better at all stages of the experiment. When the percentage of urgent tasks is high in the system (70--90\%), the Best Probability outperforms MECT and Maximum Certainty by 5-10\%, as well as the No Redirection by 10-20\%. 
As the percent of urgent tasks begins to dispel, the improvement seems to decrease a little. It is due to the fact that the system becomes less oversubscribed. Thus, we can conclude, that our proposed model offers significant performance improvement when the system is highly congested with delay sensitive tasks. 

\subsubsection{Single Base Station}
In this experiment, we aim to evaluate the performance of our model with respect to a single \bs~not considering the federation. Just as in the first experiment, we evaluate the system under an increasing oversubscription level of the system as a federation. We start initially with 2,000 vehicles per hour and increase the traffic by 1,000 vehicles until we reach  7,000 vehicles. In every trial, the simulation is run for 20 times for each heuristic. Nevertheless, unlike the first experiment, here we don't account for the overall percent of tasks that miss their deadline within the system. Instead, we evaluate the number of tasks allocated at a certain \bs~that miss their deadline. 

\begin{figure}[h!]
    \centering    
    \includegraphics[scale=0.45]{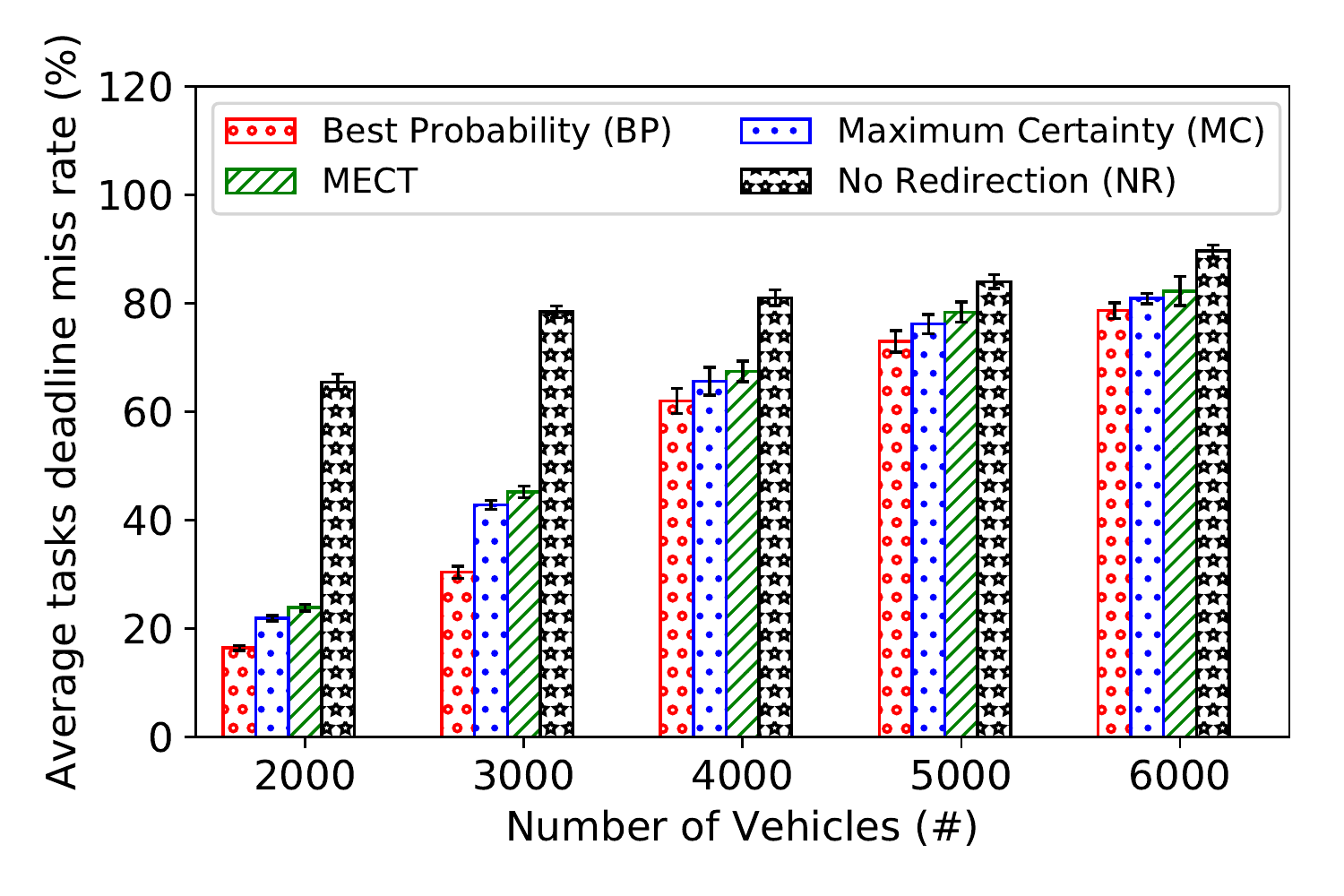}
    \caption{Average deadline miss rate while increasing the oversubscription level and considering only Base stations that miss the deadlines.}
    \label{fig:networkdelay}
\end{figure}
Figure~\ref{fig:networkdelay} shows a significant performance improvement introduced by our heuristic. The Best Probability performs consistently better then all other heuristics evaluated in the experiment. For the medium level of oversubscription (\ie \ 2,000--3,000 vehicles), Best Probability heuristic shows an improvement of approximately 45\% compared to No Redirection heuristic. Also, it outperforms MECT and Maximum Certainty heuristics by approximately 2-7\% at all stages of the experiment.


\section{Related Work}\label{relatedwork}
Efficient resource allocation that can decrease the deadline miss rate is the major challenge for V2I systems. Oversubscription situations make this challenge more complex. The robustness and QoS can deteriorate due to the lack of efficient resource allocation in a V2I system. There are several resource allocation models proposed in the literature. Jun Li et al.\cite{li2017resource} propose a local fog resource management model in Fog Enhanced Radio Access Network based on V2X environment. The core concept is to improve QoS of each \bs~for real-time vehicular services. The model considers service migration from one fog node to another based on reserved resource availability. Authors propose two resource management schemes that prioritize real-time vehicular services. Earlier versions of fog/edge computing were known as hybrid clouds~\cite{salehi10}.

Ali et al. in \cite{ali2011co} propose a multiple RSU (Road Side Unit, e.g., \bs) scheduling to provide cooperative data access to RSUs in vehicular ad-hoc networks. They categorize requests or tasks into two types (delay sensitive and delay tolerant) according to the task's data size.  In oversubscription situations, authors propose to transfer delay tolerant requests to the neighboring RSUs with a lower workload. In contrast we have utilize probabilistic model to find the suitable \bs~that increase the tasks' robustness. In \cite{liu2010rsu}  Liu and Lee suggest an RSU-based data dissemination framework to address challenges in vehicular networks. The system aims to efficiently utilize available bandwidth for both the safety-critical and the non-safety-critical services. An analytical model is proposed to investigate the system performance in terms of providing data services with delay constraints. Adachi et al. in \cite{adachi2016cloud} proposed a hybrid approach, where vehicular content (\eg \ short video clips, sensor data) is shared  among vehicles using V2V and V2I communication via cellular network. The main objective of this paper is to reduce high-cost V2I due to traffic volume in wireless network.

Although various research works have been undertaken in the field of vehicular systems \cite{maeshima2007, korkmaz2006, mak2005}, they are limited to communication functionality, quality requirements from the networking perspective, and task type prioritized scheduling. As compared to the mentioned works, we incorporate both computation and communication aspects to predict the suitable \bs~for task allocation.


\section{Conclusion and Future Works}\label{sec:conclsn}
In this paper, we proposed a robust V2I system that copes with uncertainties that exist in task arrival, communication, and computation. The robustness is provided via federating \bs s in a dynamic manner. The \lb~in each \bs~is equipped with a resource allocation method that is aware of uncertainties in communication and computation of V2I systems.
The uncertainties are captured in a normal distribution and stored in well-defined data structures, known as ETC and ETT matrices. We developed a probabilistic model that predicts the chance of successful completion of a given task on different \bs s that is leveraged in every \lb. The experiments express that the proposed model can significantly improve the robustness of the system (up to 45\%) when it is heavily oversubscribed. It is noteworthy that the system does not only improve the performance of a single \bs~but it also remarkably improves the performance of the V2I system as a whole. In the future, we will extend our probabilistic theory for cases that have heterogeneity within each \bs. We also plan to study the impact of the deadline looseness on the system. 



\section*{Acknowledgments}
Portions of this research were conducted with high performance computational resources provided by Louisiana Optical Network Infrastructure~\cite{LONI} and was supported by the Louisiana Board of Regents under grant number LEQSF(2016-19)-RD-A-25. 

\linespread{0.90}
\bibliographystyle{ieeetr} 
\balance
\bibliography{references}

\end{document}